\documentclass[twocolumn,superscriptaddress,showpacs,pre,amsmath,amssymb]{revtex4}
\usepackage{epsfig}
\begin{document}

\title{Effect of nonlinear filters on detrended fluctuation analysis}

\author{Zhi~Chen}
\affiliation{Center for Polymer Studies and Department of Physics,
               Boston University, Boston, Massachusetts 02215\\}
\author{Kun~Hu}
\affiliation{Center for Polymer Studies and Department of Physics,
               Boston University, Boston, Massachusetts 02215\\}
\author{Pedro~Carpena}
\affiliation{Departamento de F\'{\i}sica Aplicada II, ETSI de 
Telecomunicaci\'on, Universidad 
de M\'alaga, Spain\\}
\author{Pedro~Bernaola-Galvan}
\affiliation{Departamento de F\'{\i}sica Aplicada II, ETSI de 
Telecomunicaci\'on, Universidad 
de M\'alaga, Spain\\}
\author{H.~Eugene~Stanley}
\affiliation{Center for Polymer Studies and Department of Physics,
               Boston University, Boston, Massachusetts 02215\\}
\author{Plamen~Ch.~Ivanov}
\affiliation{Center for Polymer Studies and Department of Physics,
               Boston University, Boston, Massachusetts 02215\\}

\date{\today}
\pacs{05.40.-a}

\begin{abstract}

When investigating the dynamical properties of complex
multiple-component physical and physiological systems, it is often the
case that the measurable system's output does not directly represent
the quantity we want to probe in order to understand the underlying
mechanisms. Instead, the output signal is often a linear or nonlinear
function of the quantity of interest. Here, we investigate how various
linear and nonlinear transformations affect the correlation and
scaling properties of a signal, using the detrended fluctuation
analysis (DFA) which has been shown to accurately quantify power-law
correlations in nonstationary signals. Specifically, we study the
effect of three types of transforms: (i) linear ($y_i=ax_i+b$); (ii)
nonlinear polynomial ($y_i=ax_i^k$); and (iii) nonlinear logarithmic
[$y_i=\mbox{log}(x_i+\Delta)$] filters. We compare the correlation and
scaling properties of signals before and after the transform. We find
that linear filters do not change the correlation properties, while
the effect of nonlinear polynomial and logarithmic filters strongly
depends on (a) the strength of correlations in the original signal,
(b) the power $k$ of the polynomial filter, and (c) the offset
$\Delta$ in the logarithmic filter. We further apply the DFA method to
investigate the ``apparent'' scaling of three analytic functions: (i)
exponential [$\mbox{exp}(\pm x+a)$], (ii) logarithmic
[$\mbox{log}(x+a)$], and (iii) power law [$(x+a)^{\lambda}$], which
are often encountered as trends in physical and biological processes.
While these three functions have different characteristics, we find
that there is a broad range of values for parameter $a$ common for all three
functions, where the slope of the DFA curves is identical. We further
note that the DFA results obtained for a class of other analytic
functions can be reduced to these three typical cases.  We
systematically test the performance of the DFA method when 
estimating long-range power-law correlations in the output signals for
different parameter values in the three types of filters and the three
analytic functions we consider.

\end{abstract}
\maketitle
%\begin{multicols}{2}
\section{Introduction} \label{secintr}

Many physical and biological systems under multi-component control
mechanisms exhibit scale-invariant features characterized by
long-range power-law correlations in their output. These scaling
features are often difficult to quantify due to the presence of
erratic fluctuations, heterogeneity, and nonstationarity embedded in
the output signals. This problem becomes even more difficult in
certain cases: (i) when we cannot probe directly the quantity of
interest in experimental settings, i.e., the measurable output signal
is a linear or nonlinear function of the quantity of interest; (ii)
when measuring devices impose a linear or nonlinear filter on the
system's output; (iii) when we are interested not in the output signal
but in a specific component of it, which is obtained through a
nonlinear transform (e.g., the magnitude or the sign of the
fluctuations in the signal); (iv) when comparing the dynamics of
different systems by applying nonlinear transforms to their output
signals; or (v) when pre-processing the output signal by means of
linear or nonlinear filters before the actual analysis. Thus, to
understand the intrinsic dynamics of a system, in such cases it is
important to correctly analyze and interpret the dynamical patterns in
the system's output.

Conventional two-point correlation, power spectrum, and Hurst analysis
 methods are not suited for nonstationary signals, the statistical
 properties of which change with time~\cite{non,hurst1,mandelbrot1}.
 To address this problem, detrended fluctuation analysis (DFA) method
 was developed to accurately quantify long-range correlations embedded
 in a nonstationary time series \cite{CKDFA1,taqqu95}. This method
 provides a single quantitative parameter --- the scaling exponent
 $\alpha$ --- to quantify the scale-invariant properties of a signal.
One advantage of the DFA method is that it allows the detection of
long-range power-law correlations in noisy signals with embedded
polynomial trends that can mask the true correlations in the
fluctuations of a signal. Recent comparative studies have demonstrated
that the DFA method outperforms conventional techniques in accurately
quantifying correlation properties over a wide range of
scales~\cite{rmsCK,
%SVDFA1,
SMDFA1,ViswanathaPRE97,kunpre2001,janphysica2001}. The
DFA method has been widely applied to
DNA~\cite{CKDFA1,rmsCK,
%SVDFA1,
SMDFA1,mantegnaprl1994,
%CKfractal,solomdnafractal1995,
mantegnaprl1996,
%Buldyrev,
CarpenaNature02},
cardiac
dynamics~\cite{%iyengaramjphsiolreg,
HOcirc1997,
%barbiheartchaos1998,
plamenuropl1999,Pikkujamsaheartcir1999,solomrev1999,plameneuph1998,Genephsa1999,
%ashkenazyheartfrac1999,makikallioheartamjcardiol1999,crossoverCK,
Absil1999,solomphsa1999,toweillheartmed2000,bundesleep2000,Laitio2000,
%yose2000,
Yosef2001,plamenchaos2001,janpresleep02,Karasik,Echeverria,KantelhardtEurophy03},
human electroencephalographic (EEG) fluctuations~\cite{Robinson03},
human motor activity~\cite{huphysica04} and
gait~\cite{hos,Hausdorff01,Ashkenazy02,Scafetta03,West03},
meteorology~\cite{Ivanovameteo1999_12,Ivanova03}, climate temperature
fluctuations~\cite{Bundeatm,
%bundetem,
talknertem2000,Kiraly02,Eichner03,Fraedrich,Pattantyus04},
river flow and discharge~\cite{Montanari2000,Matsoukas2000}, 
%,Livina03},
electric signals~\cite{Siwy02,Varotsos1,Varotsos2}, stellar x-ray
binary systems~\cite{Moret}, neural receptors in biological
systems~\cite{bahareuph2001}, music~\cite{Jennings2004}, and
economics~\cite{%Liu97,vandewallephsa1997,
vandewallepre1998,Liu99,janosiecopha1999,ausloosphsa1999_12,robertoecopha1999,
%Vandewalle1999,
grau-carles2000,%ausloosphsa2000_9,ausloosphsa2000_10,
ausloospre2001,ausloosIntJModPhys2001}.
In many of these applications the main problem is to differentiate
scaling features in a system's output which are inherent to the
underlying dynamics, from the scaling features which are an artifact
of nonstationarities or different types of transforms and filters.

In two previous studies we have examined how different types of
nonstationarities such as superposed sinusoidal and power-law trends,
random spikes, cut-out segments, and patches with different local
behavior affect the long-range correlation properties of
signals~\cite{kunpre2001,zhipre2002}. Here we use the DFA method to
investigate how the scaling properties of noisy correlated signals
change under linear and nonlinear transforms. Further, (i) we test to
see under what types of transforms (filters) it is possible to derive
information about the scaling properties of the signal of interest
before the transformation, provided we know the correlation behavior
of the transformed (filtered) signal, and (ii) we probe the
``apparent'' scaling of three common transformation functions after 
applying the DFA method --- exponential, logarithmic and
polynomial. We also evaluate the limitations of the DFA method under
linear and nonlinear transforms. Specifically, we consider the following:

\medskip

\noindent (1) {\it Correlation properties of signals after transforms
of the type}: $\{x_i\} \Longrightarrow \{f(x_i)\}$, where $\{x_i\}$ is
a stationary signal with {\it a priori} known correlation properties.

(i) {\it Linear transform}: $\{x_i\} \Longrightarrow
\{ax_i+b\}$. Transforms of this type are often encountered in physical
systems. For example: (a) from the fluctuations in the acceleration of
a particle (measurable quantity), one can derive information about how
the force (quantity of interest) acting on this particle changes in
time without directly measuring the force: $\{a(t_i)\} \Longrightarrow
\{F(t_i)=ma(t_i)\}$; (b) in pnp-transistors a difficult to directly
measure base (input) current $I_B$ (quantity of interest) is amplified
hundreds of times, so that small fluctuations in $I_B$ may lead to
significant (and measurable) changes in the collector (output) signal
$I_C$ (measurable quantity): $\{I_C(t_i)\} \Longrightarrow
\{I_B(t_i)=I_C(t_i)/\beta\}$, and (c) changes in the volume $V$
(quantity of interest) of an ideal gas can be determined from
fluctuations in the temperature (measurable quantity) provided the
pressure is kept constant: $\{T(t_i)\} \Longrightarrow
\{V(t_i)=\frac{nR}{P}T(t_i)\}$.

(ii) {\it Nonlinear polynomial transform}: $\{x_i\} \Longrightarrow
\{ax_i^k\}$, where $k\ne 1$ and takes on positive integer values. For example:
(a) from fluctuations in the current $I$ (measurable quantity) one can
extract information about the behavior of the power lost as heat $P$
(quantity of interest) in a resistor: $\{I(t_i)\} \Longrightarrow
\{P(t_i)=RI^2(t_i)\}$; (b) measuring the temperature $T$ fluctuations
of a radiating body the Stefan's law defines the power emitted per
unit area: $\{T_i\} \Longrightarrow \{\epsilon_i=\sigma T_i^4\}$.
Further, linear and nonlinear polynomial filters are also used to
renormalize data series representing an identical quantity measured in
different systems before performing correlation analysis, e.g., (i)
normalizing heart rate recordings from different subjects to zero mean
and unit standard deviation (linear filters), or (ii) extracting the
absolute value (nonlinear filter) of the heartbeat fluctuations in
datasets obtained from different subjects~\cite{Yosef2001}.

In this study we consider two examples of nonlinear polynomial filters
--- quadratic and cubic filters --- which represent the class of
polynomial filters with even and odd powers, and we investigate how
these filters change the correlation properties of signals. Since
polynomial filters with even power wipe out the sign information in a
signal, we expect quadratic and cubic filters to have a different
effect. A recent study by Y.~Ashkenazy {\it et al.}~\cite{Yosef2001}
shows that the magnitude of a signal (without sign information)
exhibits different correlation properties from that of the original
signal. Thus it is necessary to investigate how quadratic and cubic
filters change the scaling properties of correlated signals.

(iii) {\it Logarithmic filter}: $\{x_i\} \Longrightarrow
\{\mbox{log}(x_i+\Delta)\}$, is also widely used in renormalizing
datasets obtained from different sources before comparative
analysis. For example, to compare the dynamics of price fluctuations
$X(i)$ of different company stocks, which may have a different
average price, one often first obtains the relative price returns
$R(i)=\mbox{log}[X(i+1)/X(i)]$ before performing correlation
analysis~\cite{Liu99,economicsbook_gene}. It is assumed that upon taking
the returns one does not alter the information contained in the
original signal. To test this assumption we compare the correlation
properties of the signal before and after a logarithmic filter.

\noindent (2) {\it Correlation properties of transformation functions}

When analyzing the correlation properties of a signal after a given
transform, it may be valuable to know what is the DFA result for the
transformation function itself. In addition, it is often the case that
noisy signals are superposed on trends which can be approximated by a
certain function. Previous studies have demonstrated that the DFA
result of a correlated signal with a superposed trend is a
superposition of the DFA result for the signal and the DFA result for
the analytic function representing the
trend~\cite{kunpre2001,zhipre2002}. Here we investigate separately the
results of the DFA for three functions which are very often encountered
in physical and biological processes: (i) {\it exponential}, (ii) {\it
logarithmic} and (iii) {\it power law}.

The layout of this paper is as follows: In Sec.~\ref{secpuren}, we
describe how we generate signals with desired long-range power-law
correlations and introduce the DFA method used to quantify
correlations in nonstationary signals. In Sec.~\ref{secfilters}, we
compare the correlation and scaling properties of signals before and
after linear and nonlinear polynomial transforms. In
Sec.~\ref{seclogsignals}, we consider the effect of nonlinear
logarithmic filter on the long-range correlation properties of
stationary signals. In Sec.~\ref{sectrends}, we investigate the
performance of the DFA method on three analytic functions ---
exponential, logarithmic, and power-law --- which are often
encountered as trends in physical and biological time series.  We
systematically examine the crossovers in the scaling behavior of
correlated signals resulting from the transforms and trends discussed
in Secs.~\ref{secfilters}-\ref{sectrends}, the conditions of existence
of these crossovers and their typical characteristics. We summarize
our findings in Sec.~\ref{Conclusion}.

\section{Methods}\label{secpuren}

We analyze two types of signals: 

\noindent (1) stochastic stationary signals \{$x_{i}$\}
($i=1,2,3,...,N_{\mbox{\scriptsize max}}$) with different type of
correlations (uncorrelated, correlated, and anti-correlated) and
surrogate signals obtained from \{$x_{i}$\} after linear and nonlinear
transforms. We use an algorithm based on the Fourier transform to generate
stationary signals \{$x_{i}$\} with long-range power-law correlations
as described in~\cite{CKfourier,MFFM,zhipre2002}. The generated
signals \{$x_{i}$\} have zero mean and unit standard deviation. 

\noindent(2) Exponential, logarithmic, and power-law functions which often
represent transformations or trends in physical and biological data.

We use the detrended fluctuation analysis (DFA)
method~\cite{rmsCK,
%SVDFA1,
SMDFA1} to quantify the correlation and
scaling properties of these signals. The DFA method is described in
detail elsewhere~\cite{kunpre2001,zhipre2002}. Briefly, it involves the
following steps: (i) we integrate the signal after subtracting 
the global average; (ii) we then divide the time series into boxes of
length $n$ and perform, in each box, a least-square polynomial fit of
order $\ell$ to the integrated signal to remove the local
trend in each box; (iii) in each box we calculate the root-mean-square
fluctuation function $F(n)$ quantifying the fluctuations of the
integrated signal along the local trend; (iv) we repeat this procedure
for different box sizes (time scales) $n$.
  
A power-law relation between the average root-mean-square fluctuation
function $F(n)$ and the box size $n$ indicates the presence of
scaling: $F(n) \sim n^{\alpha}$. The scale $n$ for which this scaling
holds represents the length of the correlation. The fluctuations in a
signal can be characterized by the scaling exponent $\alpha$, a
self-similarity parameter which quantifies the strength of the
long-range power-law correlations in the signal. If $\alpha=0.5$,
there is no correlation and the signal is uncorrelated (white noise);
if $\alpha < 0.5$, the signal is anti-correlated; if $\alpha >0.5$,
the signal is correlated. Since we use a polynomial fit of order
$\ell$, we denote the algorithm as DFA-$\ell$.  Further, we note 
that for stationary signals \{$x_{i}$\} with long-range power-law
correlations, the value of the scaling exponent $\alpha$ is related to
the exponent $\beta$ in the power spectrum $S(f)=f^{-\beta}$ of
signals \{$x_{i}$\} by $\beta=2\alpha-1$~\cite{rmsCK%ckpeng1991
}. Since the power spectrum is the Fourier transform of the
autocorrelation function, one can find the following relationship
between the autocorrelation exponent $\gamma$ and the power spectrum
exponent $\beta$: $\gamma=1-\beta=2-2\alpha$, where $\gamma$ is
defined by the autocorrelation function $C(\tau)=\tau^{-\gamma}$ and
should satisfy $0<\gamma<1$~\cite{janphysica2001}.

The upper threshold for the value of the scaling exponent $\alpha$ is
related to the order $\ell$ of the DFA method: $\alpha \leq \ell+1$
for DFA-$\ell$~\cite{kunpre2001}. In addition, integrating the signal 
before applying the DFA method will increase the value of the scaling
exponent $\alpha$ by 1, thus the upper threshold will become 
$\alpha+1 \leq \ell+1$ for DFA-$\ell$. Therefore, after 
integrating correlated signals with the scaling exponent $\alpha>\ell$,
one needs to apply the DFA method with an order of polynomial fit higher
than $\ell$.  We also note that for anti-correlated signals, the
scaling exponent obtained from the DFA-$\ell$ method overestimates the
true correlations at small scales~\cite{kunpre2001}. To avoid this
problem, one needs first to integrate the original anti-correlated
signal and then to apply the DFA-$\ell$ method
\cite{kunpre2001,zhipre2002}. The correct scaling exponent $\alpha$
can then be obtained from $F(n)/n$ [instead of
$F(n)$]~\cite{Yosef2001,kunpre2001,zhipre2002}. For that reason we
first integrate and then apply the DFA method when considering
anti-correlated signals.

\begin{figure*}
\centerline{
\epsfysize=0.9\columnwidth{\rotatebox{-90}{\epsfbox{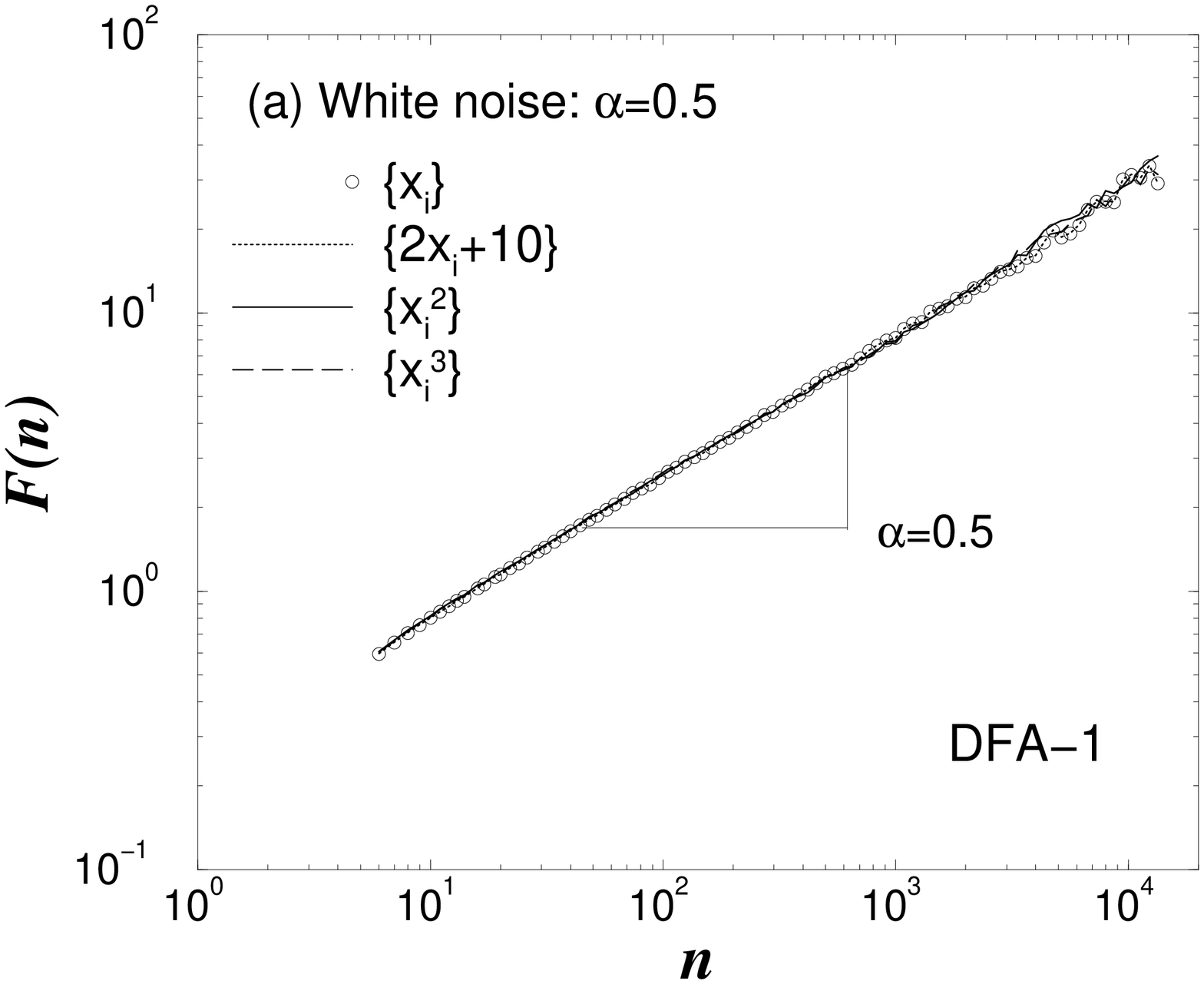}}}
\hspace{0.3cm}
%\centerline{
\epsfysize=0.9\columnwidth{\rotatebox{-90}{\epsfbox{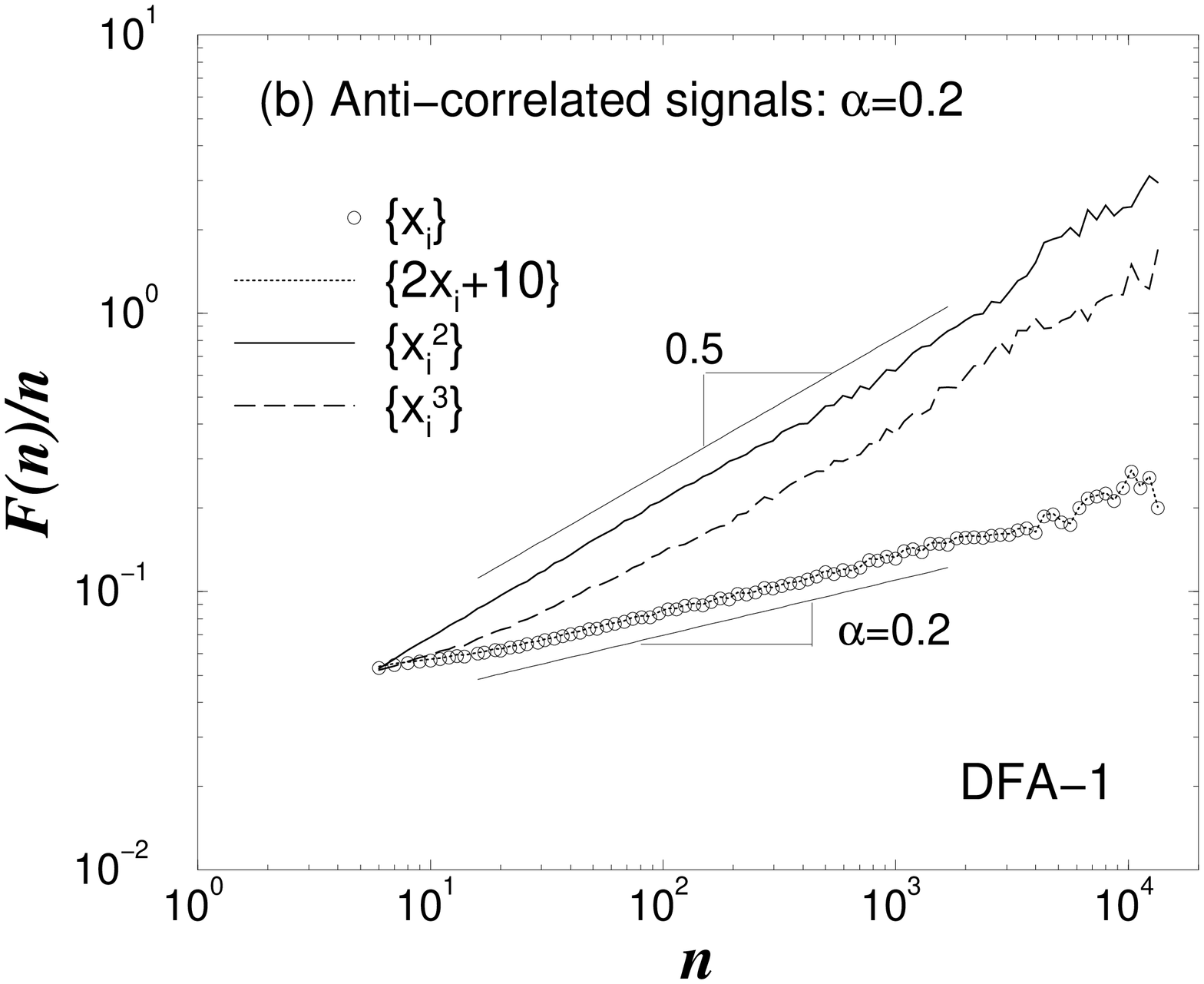}}}}
\centerline{
\epsfysize=0.89\columnwidth{\rotatebox{-90}{\epsfbox{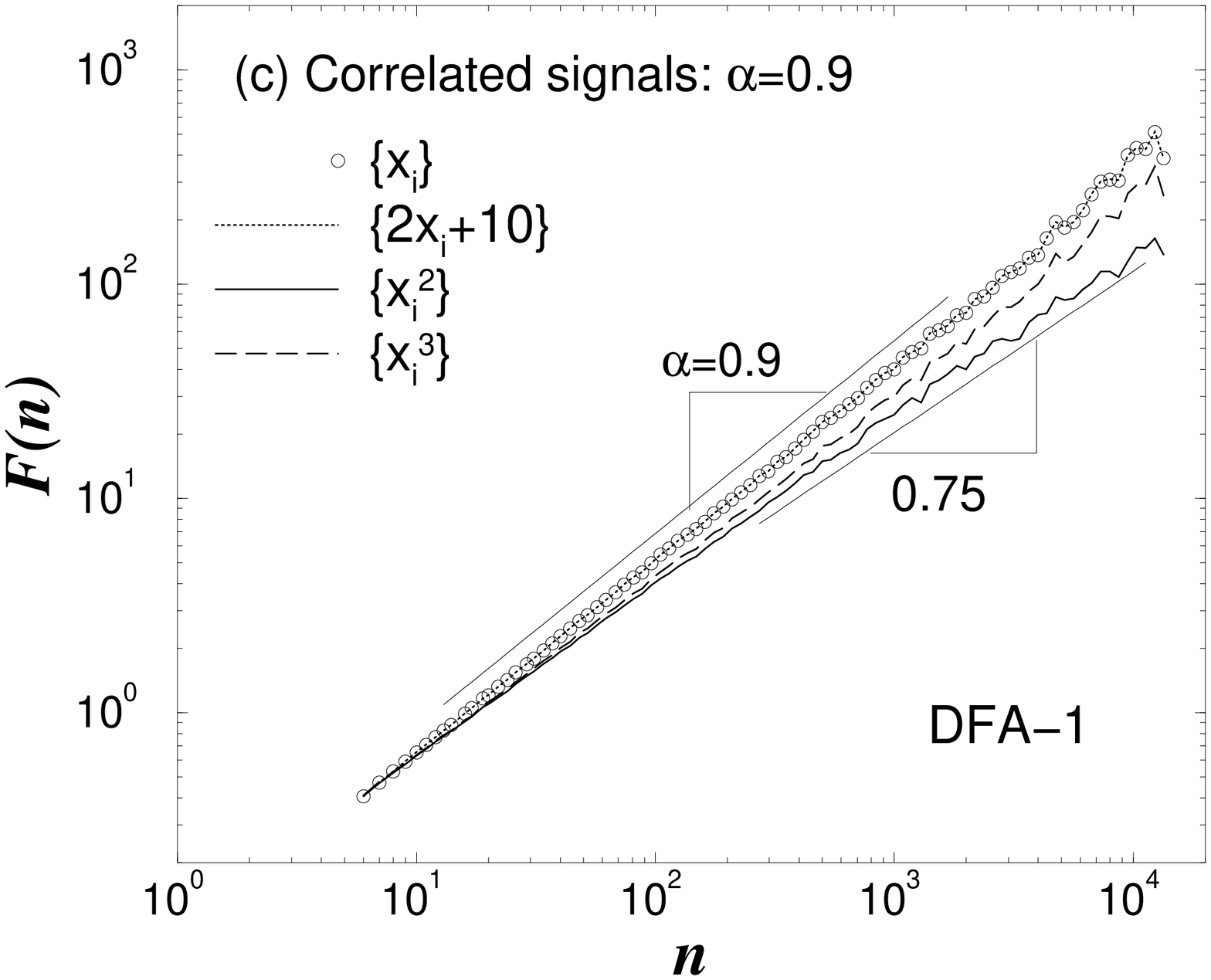}}}
\hspace{0.3cm}
%\centerline{
\epsfysize=0.9\columnwidth{\rotatebox{-90}{\epsfbox{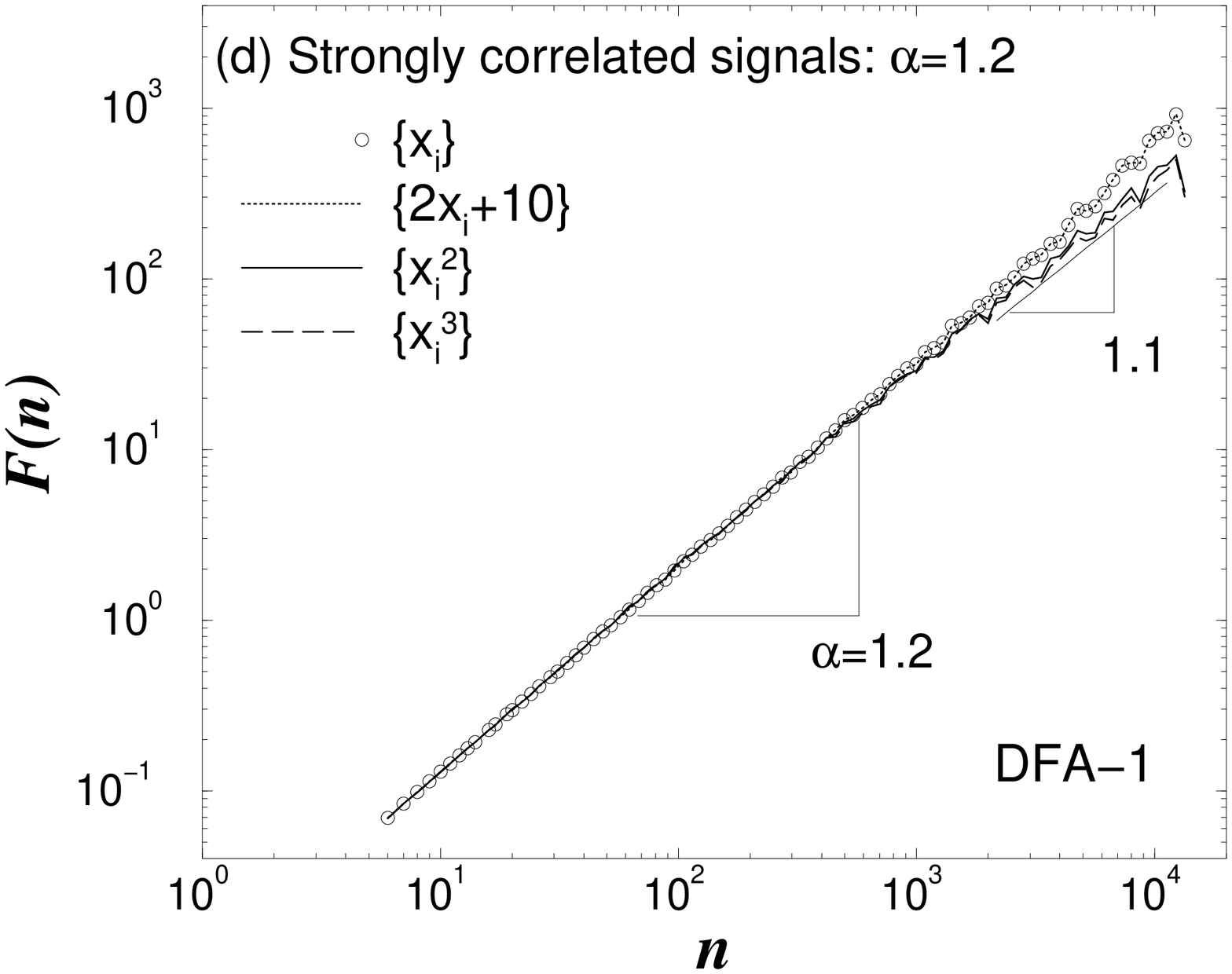}}}}

\caption{ Effects of linear, quadratic, and cubic filters on the
scaling behavior of long-range correlated stationary, Gaussian
distributed (zero mean and unit standard deviation) signals
\{$x_{i}$\}: (a) uncorrelated, (b) anti-correlated, (c) correlated,
and (d) strongly correlated. The length of each signal is
$N_{max}=2^{17}$. In our analysis we use the DFA-1 method. The curves
of the detrended fluctuation function $F(n)$ for all signals are
vertically shifted so that they start at the same value of $F(n)$ at
the minimal scale $n$. For anti-correlated signals we first integrate
and then apply the DFA-1 method to avoid overestimation of
the true correlations at small scales due to limitations of the DFA
method~(\cite{kunpre2001,zhipre2002} and Sec.~\ref{secpuren}). Our
analysis shows that after a linear filter the scaling behavior remains
unchanged. In contrast, nonlinear polynomial filters change the
scaling behavior of anti-correlated and correlated signals, leading to
crossovers at large scales.
}
\label{filters}
\end{figure*}

\section{Effects of linear and nonlinear polynomial transforms}\label{secfilters}

In this section, we study the effect of linear and nonlinear polynomial
transforms (filters) on the scaling properties of stationary signals
\{$x_{i}$\} with long-range power-law correlations. Specifically, we
consider two types of nonlinear transforms --- quadratic and cubic ---
as an example of even and odd polynomial filters. We generate the
signals \{$x_{i}$\} with linear fractal properties and with {\it a
priori} build-in correlations characterized by a DFA scaling exponent
$\alpha$~\cite{CKDFA1,kunpre2001,zhipre2002}. We compare how the
exponent $\alpha$ changes after the transform.

We first test to see if these transforms affect the properties of
uncorrelated signals (white noise). We find that the linear, quadratic, 
and cubic filters do not change the scaling properties of white noise
--- the curves of the detrended fluctuation function $F(n)$ for the
filtered signals \{$f(x_i)$\} collapse on the scaling curve of the
original signal \{$x_{i}$\}, and the scaling exponent $\alpha=0.5$
remains unchanged [Fig.~\ref{filters}(a)].

For signals with correlations we find that the linear and nonlinear
polynomial filters have a different effect. In particular, for both
correlated ($\alpha>0.5$) and anti-correlated ($\alpha<0.5$) signals
\{$x_{i}$\} we find that the scaling properties remain unchanged after
the linear filter. In contrast, the quadratic and cubic filters change the
scaling behavior of both correlated and anti-correlated signals
[Fig.~\ref{filters}(b-d)]. Specifically, for {\it anti-correlated}
signals, we find that: (i) after the quadratic filter the scaling behavior
is dramatically changed to uncorrelated (random) behavior with
$\alpha=0.5$ at all scales; (ii) after the cubic filter the scaling
(correlation) function $F(n)$ of anti-correlated signals is also
changed and exhibits a crossover from anti-correlated behavior at
small scales to uncorrelated behavior at larger scales
[Fig.~\ref{filters}(b)]. We note that the quadratic filter removes the
sign information in a signal, thus completely eliminating the
anti-correlations in a signal. In contrast, the effect of the cubic
filter is not as strong as the effect of the quadratic filters, since
a cubic filter preserves the sign information and the anti-correlations
at small scales. For {\it correlated} signals we find that after both
quadratic and cubic filters, the scaling behavior is unchanged at small
and intermediate scales. At large scales we observe a crossover to
weaker correlations which is less pronounced when increasing the
strength of the correlations (higher values of $\alpha$) in the signal
\{$x_{i}$\} [Fig.~\ref{filters}(c-d)]. For signals with very strong
correlations ($\alpha>1$), we find that the scaling behavior remains
almost unchanged after nonlinear polynomial filters. We also find that
the quadratic filter leads to a more pronounced crossover at large
scales compared to the cubic filter for all positively correlated
signals.

\section{logarithmic filter}\label{seclogsignals}

\begin{figure*}  
\centerline{
\epsfysize=0.9\columnwidth{\rotatebox{-90}{\epsfbox{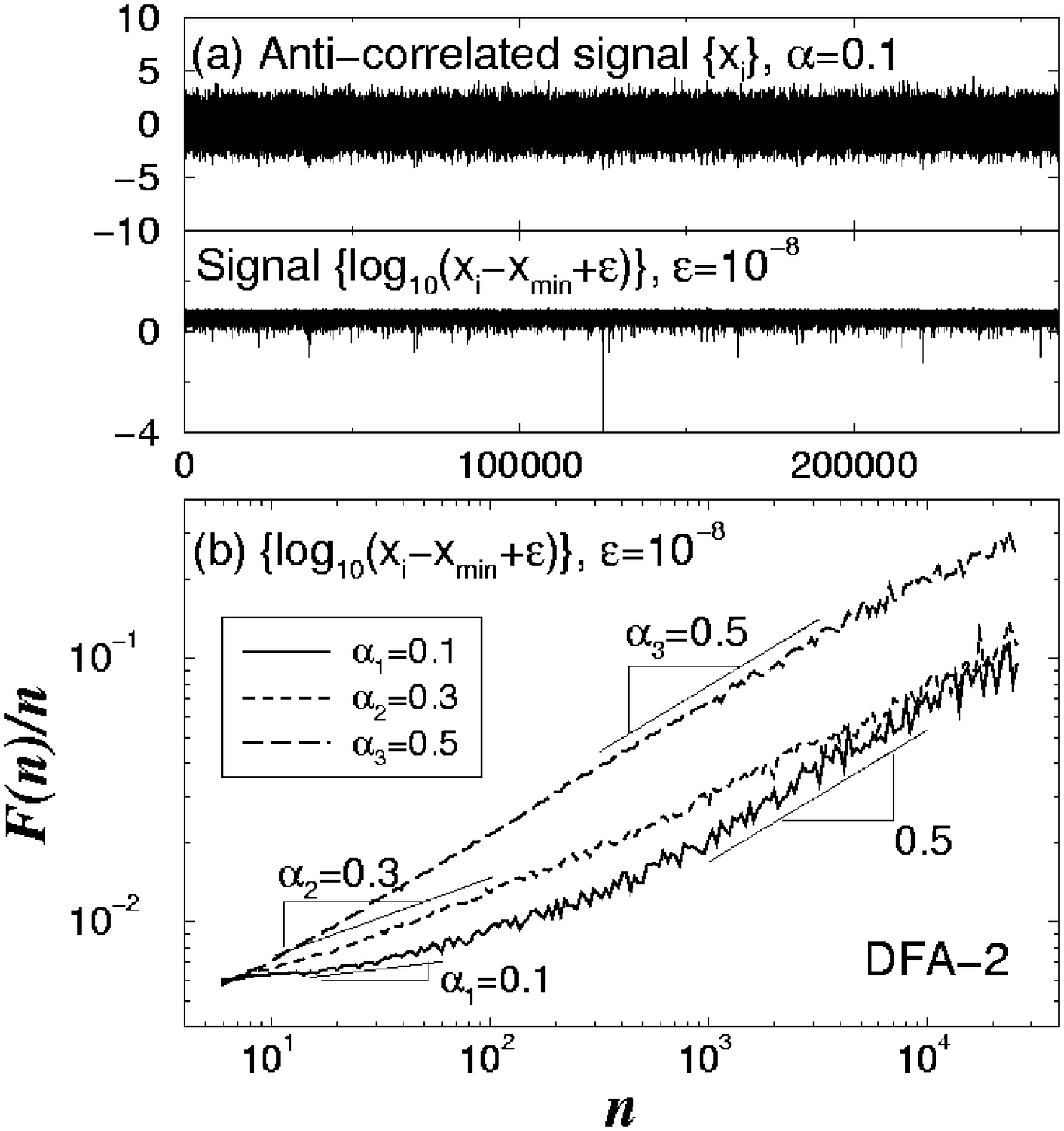}}}
\hspace{0.5cm}
\epsfysize=0.9\columnwidth{\rotatebox{-90}{\epsfbox{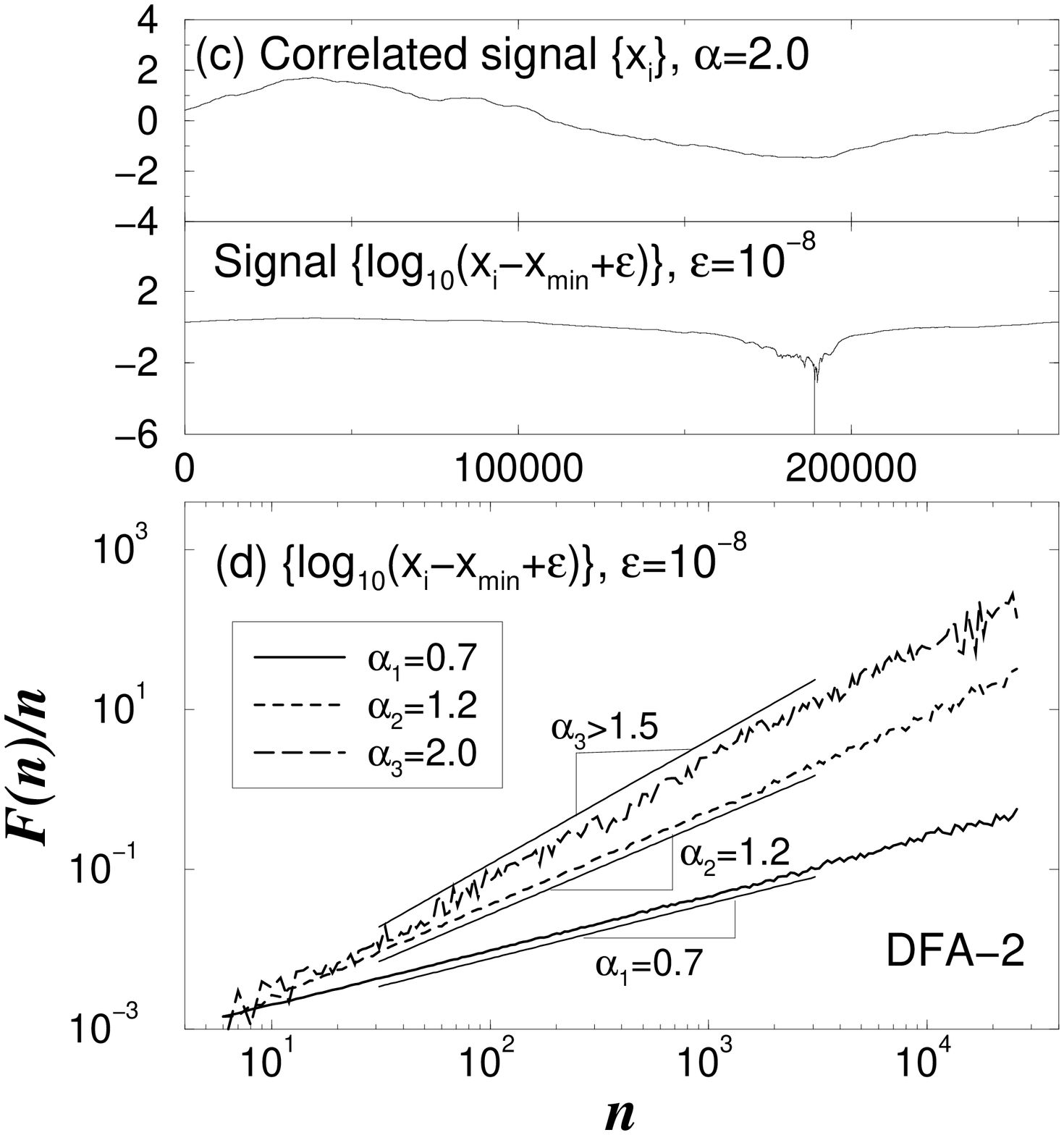}}}}
\caption{Effects of the nonlinear logarithmic filter
$\{\mbox{log}_{10}$($x_{i}- x_{min}+\epsilon)\}$ on the scaling
behavior of stationary correlated signals \{$x_{i}$\}, where $x_{min}$
is the minimal value in the original signal $\{x_i\}$ and $\epsilon$
is a positive constant. The original signals $\{x_i\}$ have zero mean,
unit standard deviation, and length $N_{max}=2^{18}$.  (a) Original
strongly anti-correlated signal \{$x_{i}$\} with DFA correlation
exponent $\alpha=0.1$ and the corresponding signal after logarithmic
filter.  (b) DFA scaling curves $F(n)$ for anti-correlated signals and
white noise after the logarithmic filter show a crossover to ``white
noise'' behavior (i.e., slope$=0.5$) at large scales.  To obtain more
accurate scaling, we first integrate the signal
$\{\mbox{log}_{10}$($x_{i}- x_{min}+\epsilon)\}$ and then apply DFA-2
method (see Sec.~\ref{secpuren}).  (c) Original strongly correlated
signal \{$x_{i}$\} with the DFA correlation exponent $\alpha=2$ and the
corresponding signal after logarithmic filter.  (d) DFA scaling curves
for correlated signals \{$x_{i}$\} after the logarithmic filter.  We find
that the logarithmic filter does not change the correlation properties
of signals with certain positive correlations (e.g., $\alpha=0.7$ and
$\alpha=1.2$), though it weakens the correlations in signals with very
strong positive correlations (e.g., $\alpha=2$).}
\label{logsg1} 
\end{figure*}

\begin{figure*}  
\centerline{
\epsfysize=0.9\columnwidth{\rotatebox{-90}{\epsfbox{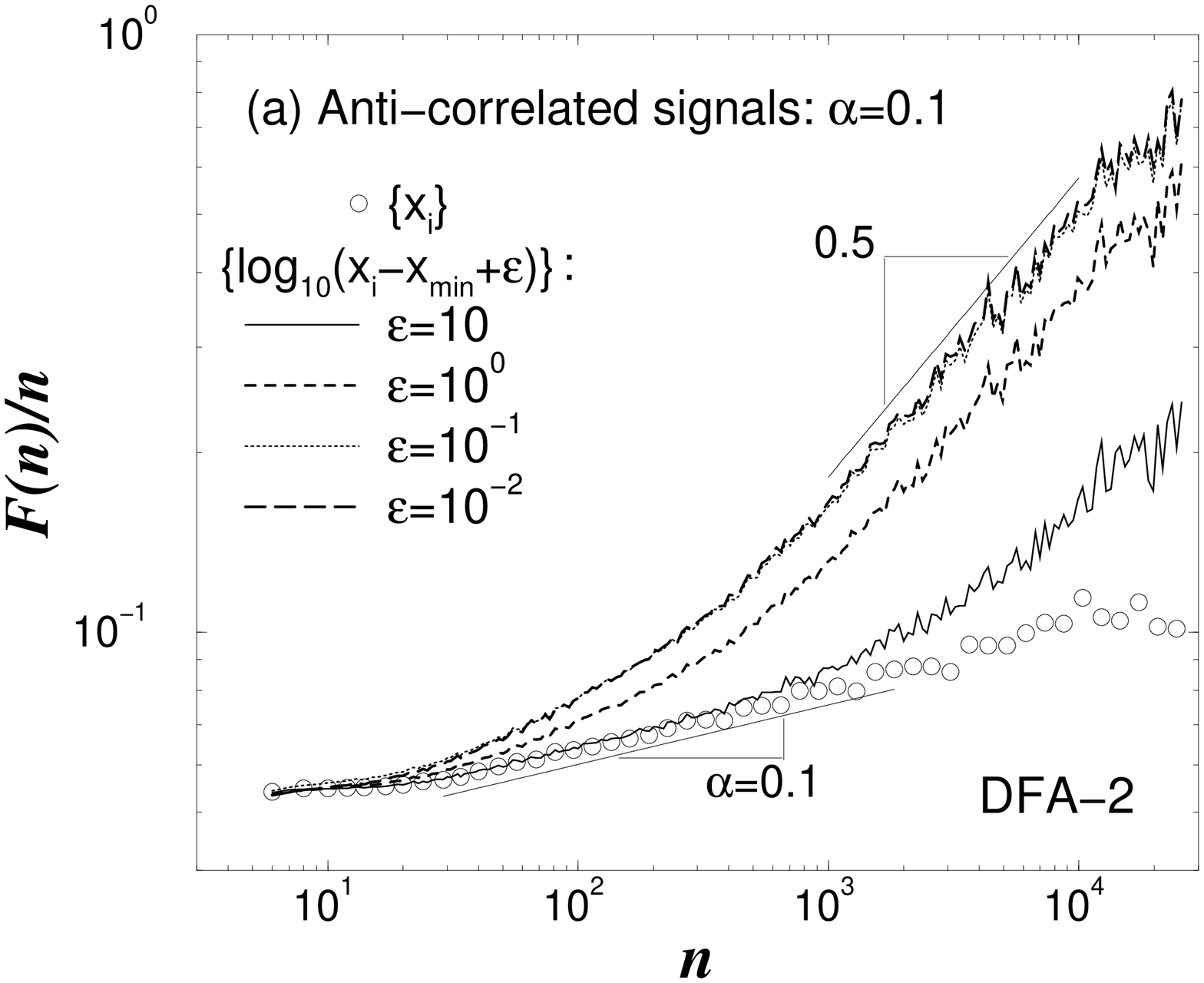}}}
\hspace{0.5cm}
\epsfysize=0.9\columnwidth{\rotatebox{-90}{\epsfbox{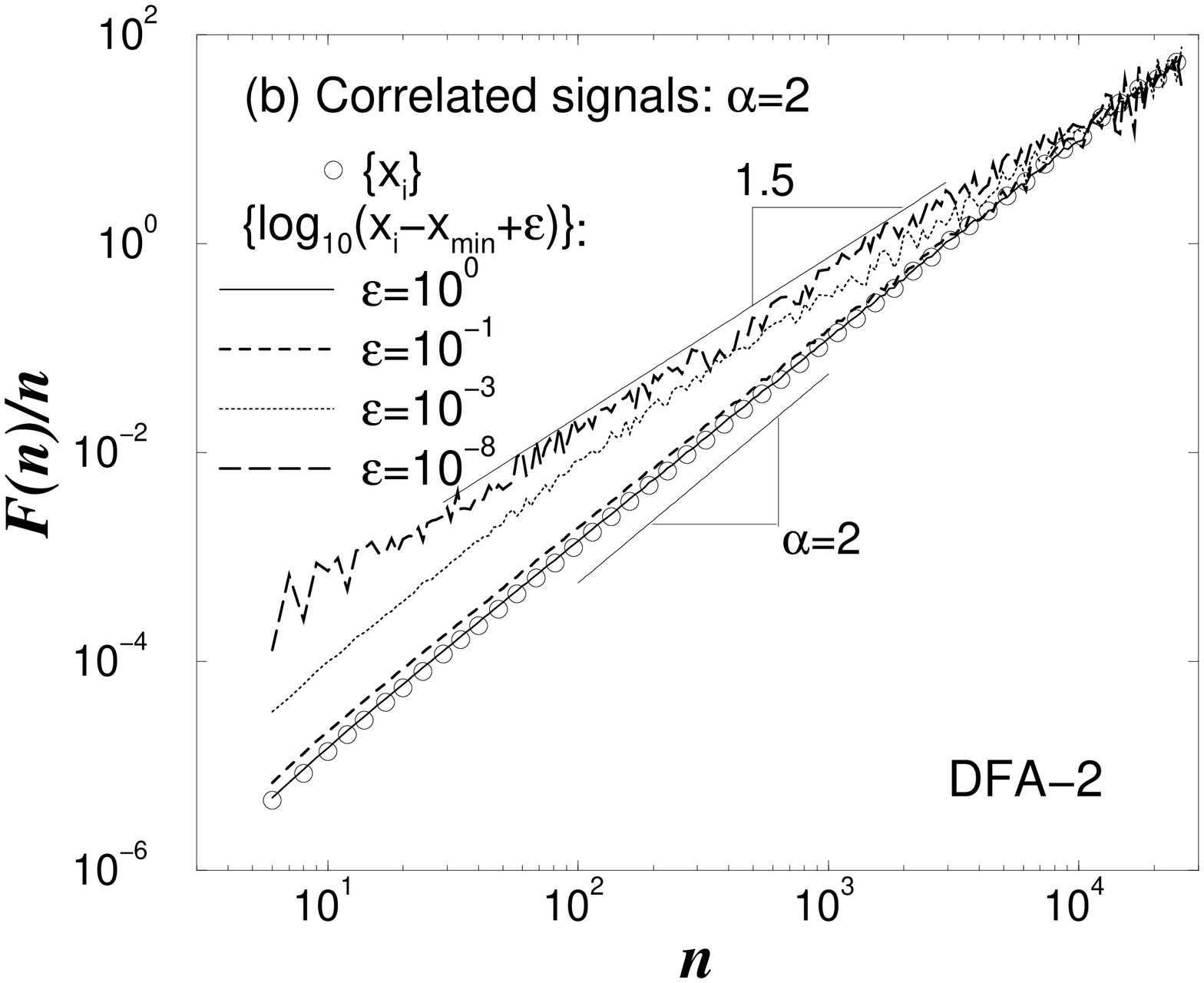}}}}
\caption{ Dependence of the effect of logarithmic filter
$\{\mbox{log}_{10}(x_{i}-x_{min}+\epsilon)\}$ on the offset parameter
$\epsilon$.  (a) Detrended fluctuation function $F(n)$ from the DFA-2
after integration of $\{\mbox{log}_{10}(x_{i}-x_{min}+\epsilon)\}$,
for an anti-correlated signal with the DFA correlation exponent $\alpha=0.1$
and varied values of $\epsilon$. We find that for smaller values of
$\epsilon$, there is a more pronounced crossover to uncorrelated
behavior with $\alpha=0.5$. (b) Detrended fluctuation function $F(n)$
from the DFA-2 after integration of
$\{\mbox{log}_{10}(x_{i}-x_{min}+\epsilon)\}$, for a signal with
strong positive correlations ($\alpha=2$) and varied values of
$\epsilon$. We find that signals with strong positive correlations are
less affected by the logarithmic filter compared to the
anti-correlated signals in (a) and that for smaller values of
$\epsilon$, there is a more pronounced crossover.
}
\label{logsg2} 
\end{figure*}

In addition to nonlinear polynomial transforms, logarithmic transforms
are often used in preprocessing procedures when there is a need to
renormalize output signals obtained from different systems before
comparing their correlation properties~\cite{%Liu97,
Liu99}. In this
section, we investigate the effect of logarithmic filters on the
scaling properties of stationary signals with long-range power-law
correlations.

We first generate stationary correlated signals \{$x_{i}$\} with a zero
mean and unit standard deviation, and with {\it a priori} known and
controlled correlation properties quantified by DFA scaling exponent
$\alpha$. To ensure that all values in the signal are positive, before
the logarithmic transform, we shift $\{x_{i}\} \Longrightarrow
\{x_i+\Delta\}$, where $\Delta=-x_{min}+\epsilon$, $x_{min}$ is the
minimal value in the series \{$x_{i}$\} and $\epsilon$ is a positive
constant. This linear transform does not alter the correlation
properties of \{$x_{i}$\}, as demonstrated in Sec.~\ref{secfilters},
Fig.~\ref{filters}. Next we integrate the signal after the logarithmic
transform $\{\mbox{log}_{10}(x_{i}-x_{min}+\epsilon)\}$ and we perform a 
DFA-2 analysis.

For uncorrelated (white noise) signals after the logarithmic filter,
we find no change in the scaling properties and the correlation
exponent remains $\alpha=0.5$ in the entire range of scales
[Fig.~\ref{logsg1}(b)]. However, we find that the scaling properties of
signals with certain degree of correlation change
significantly. Specifically, for anti-correlated signals
($\alpha<0.5$) we observe a crossover to uncorrelated (white noise)
behavior at large scales. This crossover becomes more pronounced (and
shifted to smaller scales) when increasing the strength of
anti-correlations (decreasing $\alpha$) [Fig.~\ref{logsg1}(b)]. This
crossover behavior is caused by negative spikes in the signal following 
the logarithmic transform [Fig.~\ref{logsg1}(a)]. A similar effect was
previously reported for stationary correlated signals with superposed
random spikes~\cite{zhipre2002}. For correlated signals ($\alpha>0.5$), 
we find a threshold value for the correlation exponent
$\alpha_{th}\approx 1.3$, below which the scaling properties of the
signal remain unchanged after the logarithmic filter. Above
$\alpha_{th}$ there is a reduction in the strength of the positive
correlations, i.e., the value of the estimated exponent after the
logarithmic filter is much lower compared to the correlation exponent
$\alpha$ in the original signal [Fig.~\ref{logsg1}(d)].

\begin{figure}  
\centerline{
\epsfysize=0.9\columnwidth{\rotatebox{-90}{\epsfbox{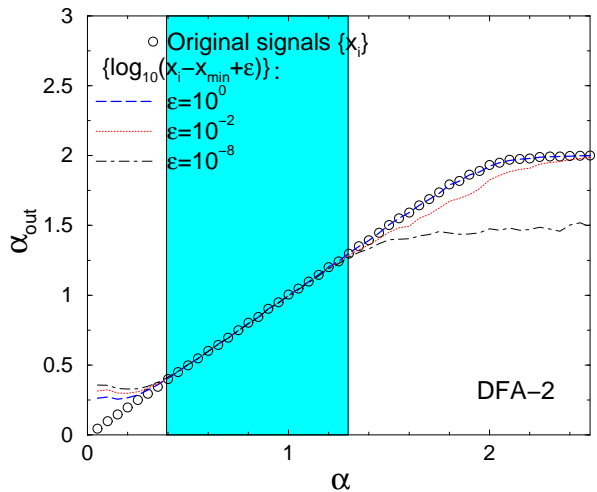}}}}
\caption{Relation between the scaling exponent $\alpha$ of the
original ``input'' stationary signals and the correlation exponent
$\alpha_{out}$ of the signals after the logarithmic filter
\{$\mbox{log} (x_{i}-x_{min}+\epsilon)$\}, where $x_{min}$ is the
minimal value in the original signal $\{x_i\}$ and $\epsilon$ is a
positive constant. $\alpha_{out}$ is obtained from the DFA-2 analysis
after integrating the signal \{$\mbox{log} (x_{i}-x_{min}+\epsilon)$\}
and fitting the detrended fluctuation function $F(n)$ in the region
$n\in [30, 3000]$. Our results show that for signals with a
correlation exponent $\alpha$ outside the shaded region, the
logarithmic filter changes the scaling behavior ($\alpha_{out} \ne
\alpha$) and this change depends on the offset parameter $\epsilon$.}
\label{logsg3} 
\end{figure}

Since the logarithmic filter is a nonlinear transform which diverges
for values of the signal \{$x_i-x_{min}+\epsilon$\} close to zero, we
next test how the scaling properties of the signal depend on the value
of the offset parameter $\epsilon$. We consider anti-correlated and
correlated signals with fixed values of $\alpha$ and varied
$\epsilon$. For strongly anti-correlated signals we find that even for large
values of $\epsilon$, there is a crossover to uncorrelated behavior
in the scaling curve $F(n)$ at large scales (note that $\epsilon$ is
the minimal value of the signal \{$x_i-x_{min}+\epsilon$\}). This
crossover shifts to smaller scales with decreasing $\epsilon$
[Fig.\ref{logsg2}(a)]. Further, we find that for decreasing $\epsilon$,
the scaling curves $F(n)$ converge to a single curve, indicating
random uncorrelated behavior in the range of large and intermediate
scales. For anti-correlated signals with $\alpha=0.1$ we find that
this convergence is reached for $\epsilon<0.1$
[Fig.\ref{logsg2}(a)]. For signals with strong positive correlations
($\alpha>\alpha_{th}$), we also observe a change in the scaling
behavior which becomes more pronounced when $\epsilon$
decreases. However, in contrast to the anti-correlated signals, the
deviation from the expected accurate scaling starts at intermediate
scales and extends to smaller scales with decreasing $\epsilon$
[Fig.\ref{logsg2}(b)]. For signals with very strong correlations,
e.g. $\alpha=2$, the deviation from the accurate scaling is observed
only for $\epsilon<0.1$, while for $\epsilon>0.1$, there is no
effect on the scaling [Fig.\ref{logsg2}(b)]. This is in contrast to the
situation observed for signals with strong anti-correlations
($\alpha=0.1$) where the logarithmic filter alters the scaling
behavior even for much larger values $\epsilon>10$
[Fig.\ref{logsg2}(a)].

Finally we study the relation between the scaling exponent $\alpha$ of
the original ``input'' signal and the estimated exponent
$\alpha_{out}$ of the ``output'' signal after the logarithmic
filter. We find that for correlated signals within given range for the
value of the scaling exponent $\alpha \in [0.4,1.3)$, there is no
change in the scaling properties after the logarithmic
transform. However, for signals with correlation exponents
$\alpha<0.4$ and $\alpha>1.3$, we find that the logarithmic transform can
dramatically change the scaling behavior and this effect also strongly
depends on the value of the offset parameter $\epsilon$
[Fig.\ref{logsg3}]. Therefore, the logarithmic filter is not recommended
for anti-correlated signals and signals with very strong positive
correlations --- applying this filter will mask the true
correlations in the original signals.

\section{results of the DFA for transformation functions}
\label{sectrends}

\begin{figure*}
\centerline{
\epsfysize=0.9\columnwidth{\rotatebox{-90}{\epsfbox{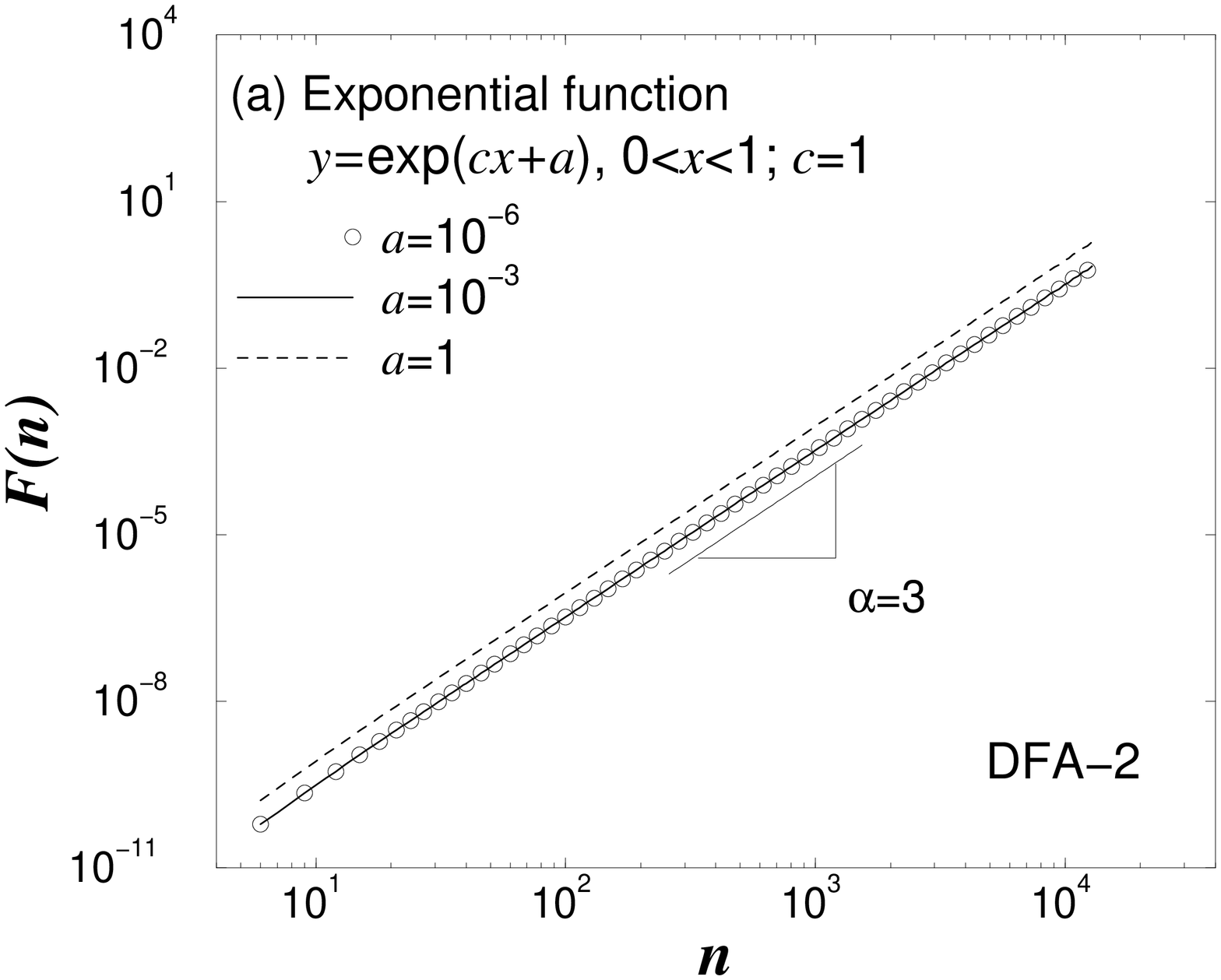}}}
\epsfysize=0.9\columnwidth{\rotatebox{-90}{\epsfbox{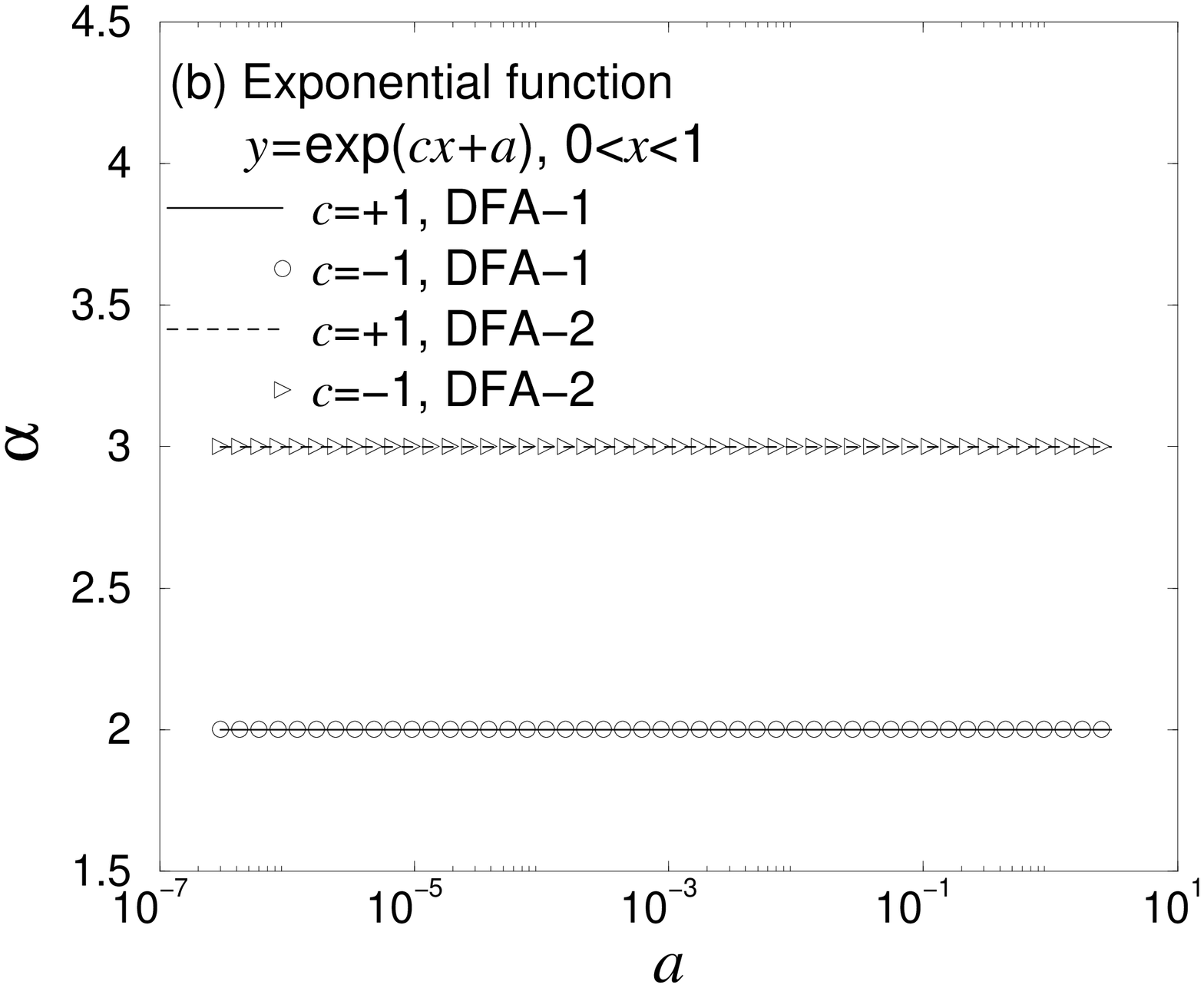}}}}
\caption{The results of the DFA method for general exponential function:
$y=\mbox{exp}(cx+a), 0<x\leq 1, x=i/N_{max},
i=1,2,...,N_{max}, N_{max}=2^{17}$, where $c=\pm1$ and offset $a$ is a positive
constant. (a) Detrended fluctuation function $F(n)$ obtained using the
DFA-2 method for different values of the offset parameter $a$. While
there is a vertical shift in $F(n)$ for different values of $a$, all
scaling curves are characterized by an identical slope $\alpha$. (b)
Dependence of the scaling exponent $\alpha$ on the parameters $a$ and
$c$. We find that for any exponential function the scaling exponent
$\alpha$ depends only on the order $\ell$ of the DFA method:
$\alpha=\ell+1$.}
\label{trends3}
\end{figure*}

In this section we investigate the scaling properties of three
functions: {\it exponential}, {\it logarithmic}, and {\it
power-law}. These functions are often used in signal processing as
transforms of various stochastic correlated signals and also appear as
trends superposed on noisy signals derived from physical and
biological systems. In previous work~\cite{kunpre2001,zhipre2002} we
have demonstrated that the scaling behavior of a correlated signal
with a superposed trend is superposition of the scaling behavior of
the correlated signal and the ``apparent'' scaling behavior obtained
from the DFA method for the analytic function representing the
trend. Therefore, understanding the results of the DFA for certain
analytic functions becomes a necessary step to quantify the scaling
behavior of system's outputs where correlated fluctuations are
superposed with different trends.

\begin{figure*}
\centerline{
\epsfysize=0.98\columnwidth{\rotatebox{-90}{\epsfbox{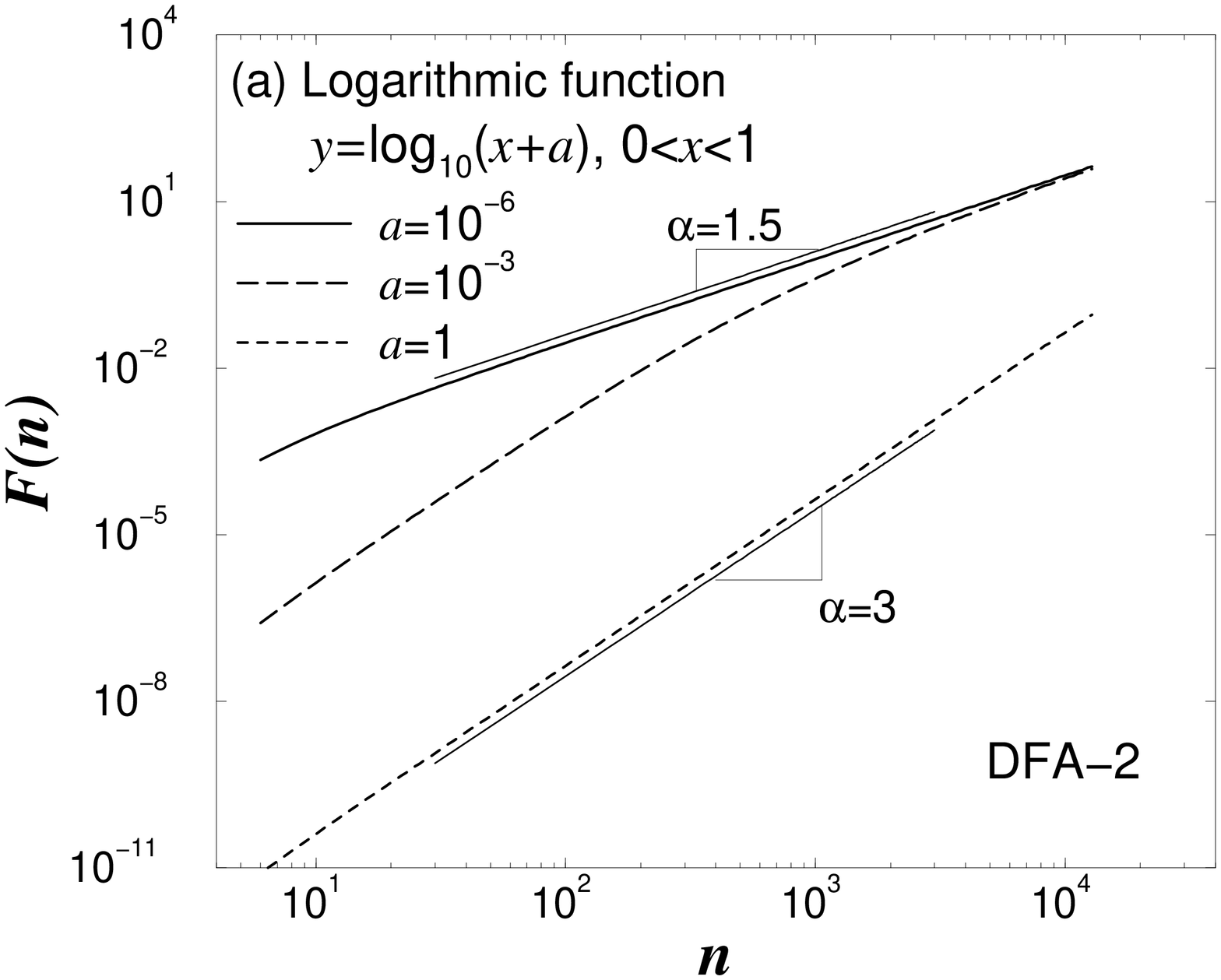}}}
\epsfysize=0.9\columnwidth{\rotatebox{-90}{\epsfbox{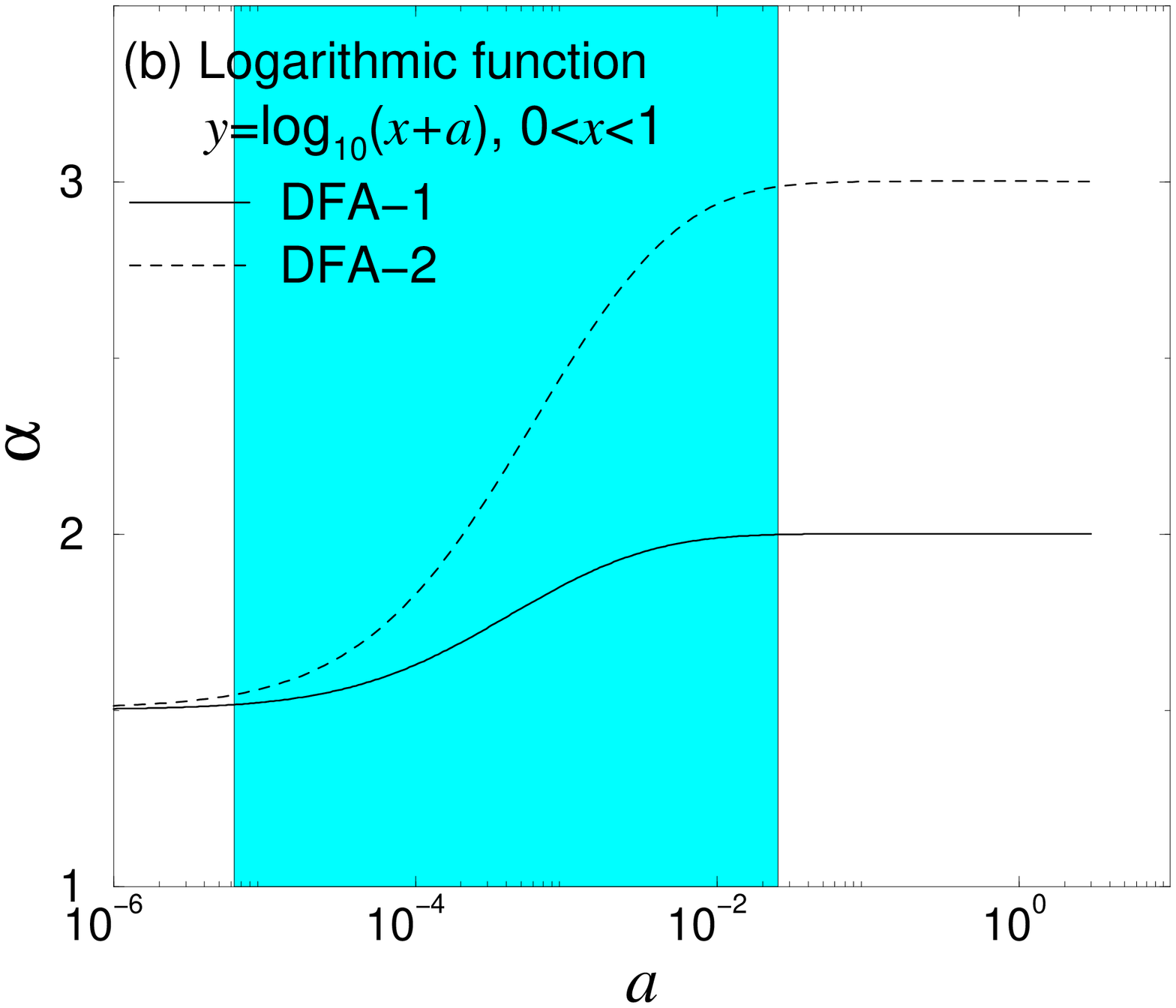}}}}
\caption{The results of the DFA method for general logarithmic function
$y=\mbox{log}_{10}(x+a), 0<x\leq 1, x=i/N_{max}, i=1,2,...,N_{max},
N_{max}=2^{17}$, where offset $a$ is a positive constant. (a)
Detrended fluctuation function $F(n)$ obtained using the DFA-2 method
for different values of the offset parameter $a$. We find that the
slope of the scaling curve (scaling exponent $\alpha$) depends on the
value of the offset $a$. (b) Dependence of the scaling exponent
$\alpha$ on the offset $a$ [fitting region for $\alpha$ is $n\in
(30,3000)$].  We observe a dramatic change from $\alpha=1.5$ at
$a\simeq0$ to $\alpha=\ell+1$ at $a>0.01$, where $\ell$ is the order
of the DFA method.}
\label{trends2}
\end{figure*}

(i) {\it We first consider the exponential function in the form}:
$y=\mbox{exp}(cx+a)$, where $0<x\leq 1$, $x=i/N_{max}$, $i=1,...,
N_{max}$, $N_{max}=2^{17}$, the parameter $c=\pm 1$, the offset
parameter $a$ is a positive constant.  We show the result of the DFA
method in Fig.~\ref{trends3}. We find that the slope of the detrended
fluctuation function $F(n)$ vs. the scale $n$ obtained from the DFA
method does not depend on the values of the parameters $c$ and $a$
(there is only a vertical shift in $F(n)$ for different values of $a$
and $c$) [Fig.~\ref{trends3}(a)]. Instead, we find that the DFA scaling
exponent $\alpha$ depends only on the order $\ell$ of polynomial fit
in the DFA method --- $\alpha=\ell+1$ --- suggesting that the results
of the DFA method do not depend on the details of the exponential
function [Fig.~\ref{trends3}(b)]. An analytic derivation for the
fluctuation function $F(n)$ and the value of the scaling exponent
$\alpha$ obtained from DFA-1 is presented in Appendix A.

\begin{figure*}
\centerline{
\epsfysize=0.94\columnwidth{\rotatebox{-90}{\epsfbox{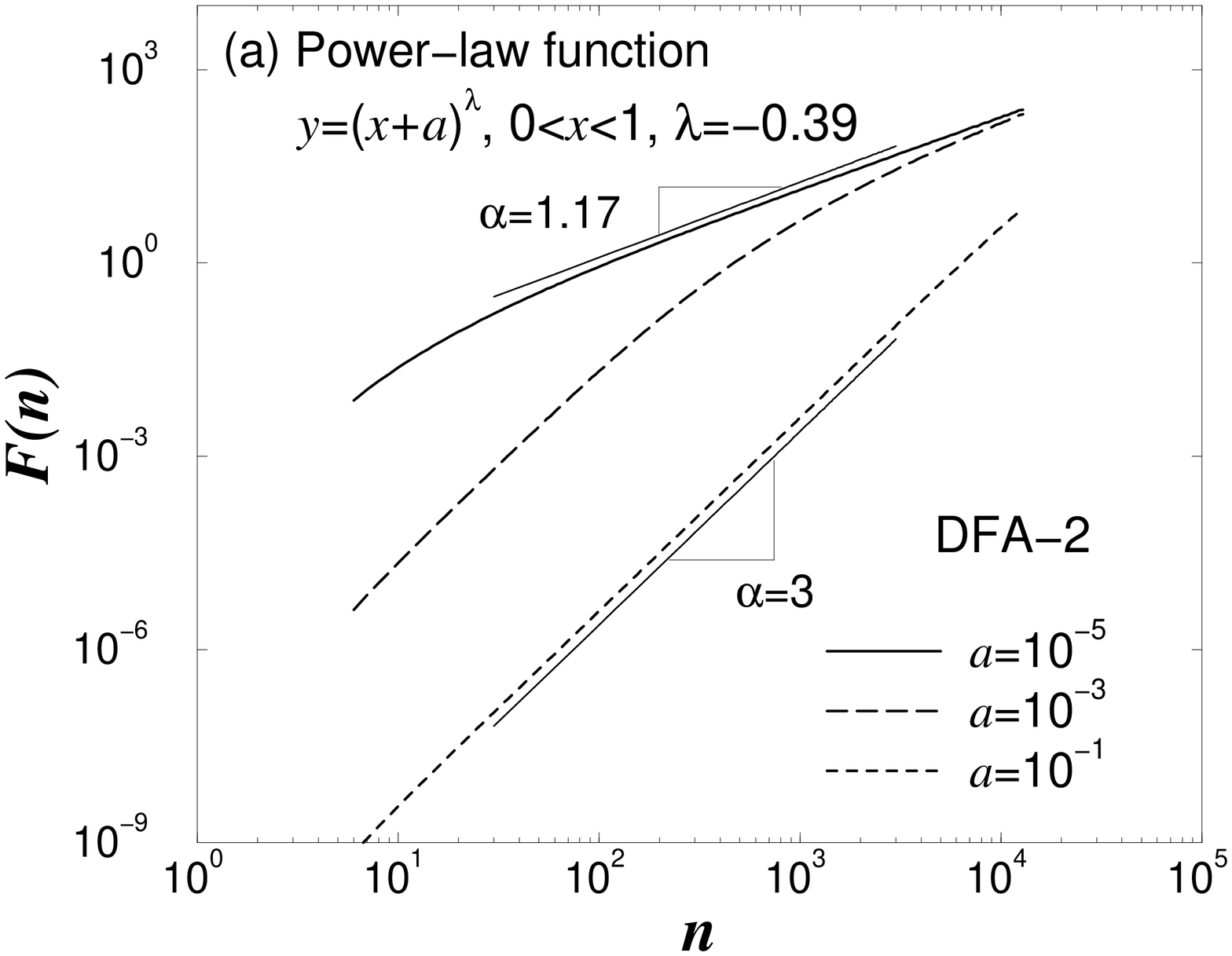}}}
\epsfysize=0.9\columnwidth{\rotatebox{-90}{\epsfbox{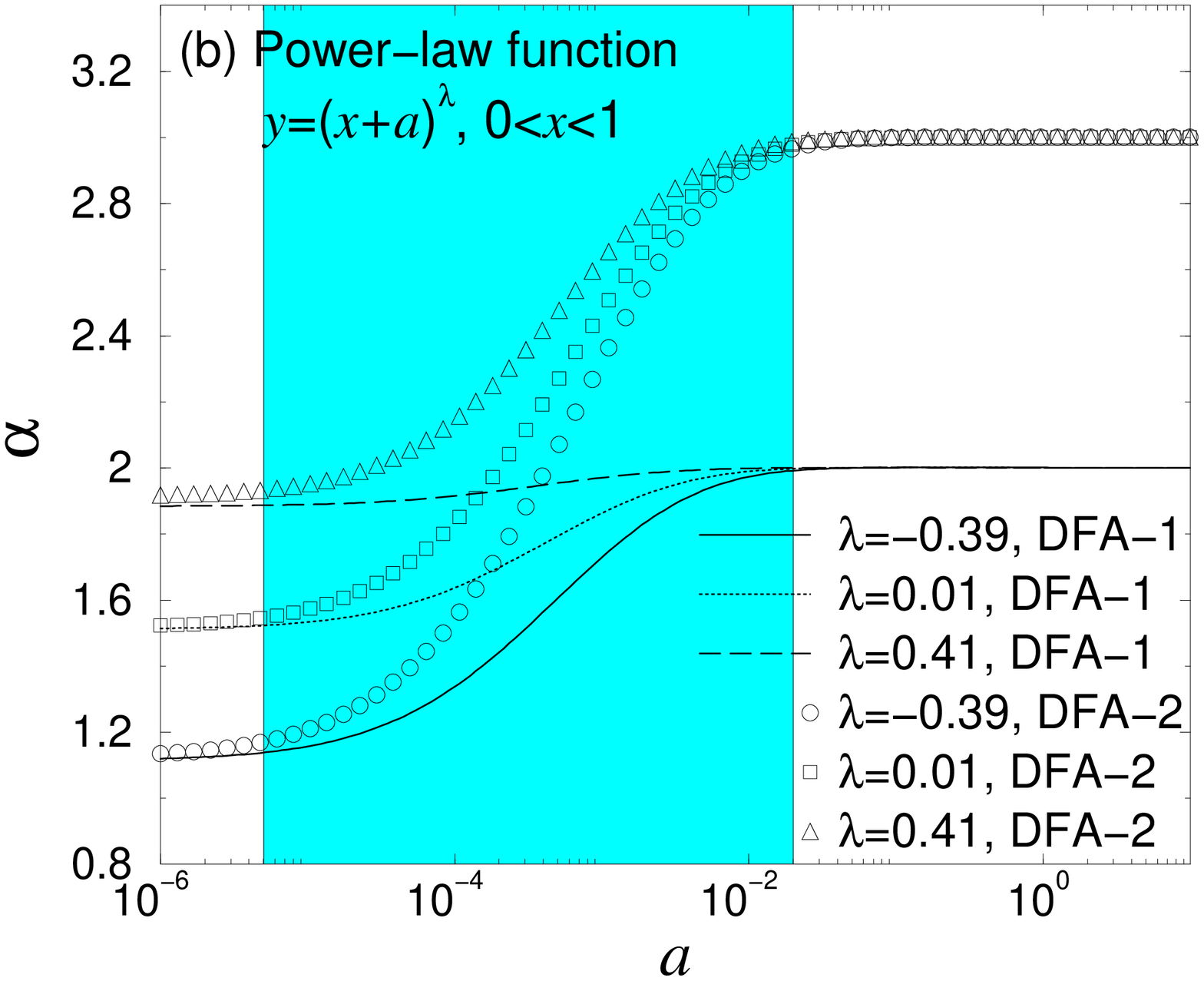}}}}
\caption{The results of the DFA method for general power-law function
$y=(x+a)^{\lambda}, 0<x\leq 1, x=i/N_{max}, i=1,2,...,N_{max}, N_{max}=2^{17}$,
where $\lambda$ is the power and the offset parameter $a$ is a
positive constant. (a) Detrended fluctuation function $F(n)$ obtained
using the DFA-2 method for fixed $\lambda=-0.39$ and different values
of the offset parameter $a$.  We find that the slope of the scaling
curve (scaling exponent $\alpha$) depends on the value of $a$.  (b)
Dependence of the scaling exponent $\alpha$ on the offset
$a$ for different values of the power $\lambda$ [fitting region for
$\alpha$ is $n\in (30,3000)$]. We observe a dramatic change from
$\alpha\simeq \lambda+1.5$ at $a\simeq 0$ for different values of
$\lambda$ to $\alpha=\ell+1$ for $a>10^{-2}$, where $\ell$ is the
order of the DFA method.}
\label{trends1}
\end{figure*}

(ii) {\it We next consider the performance of the DFA method on a logarithmic
function of the general form}: $y=\mbox{log}_{10}(x+a)$, where
$0<x\leq 1$,~$x=i/N_{max}$, $i=1,..., N_{max}$, $N_{max}=2^{17}$ and
the offset parameter $a$ is a positive constant. Specifically, we
investigate the dependence of the DFA scaling exponent $\alpha$ on the
value of the offset parameter $a$. We find that for very small values
of $a$, the DFA scaling exponent is $\alpha=1.5$. With increasing $a$,
we observe a crossover in $F(n)$ at intermediate scales $n$ --- from
$\alpha=1.5$ at large scales to $\alpha=3$ at small scales for DFA-2
[Fig.~\ref{trends2}(a)]. For larger values of $a$, we observe a scaling
behavior in $F(n)$ characterized by a single exponent $\alpha=3$ in
the entire range of scales $n$ [Fig.~\ref{trends2}(a)]. In
Fig.~\ref{trends2}(b) we present the dependence of the DFA scaling
exponent $\alpha$ [obtained in the fitting range $n \in (30,3000)$] on
the offset parameter $a$ for different DFA order $\ell$. We find that
for $a<10^{-5}$ the exponent $\alpha$ does not depend on the order
$\ell$ of the DFA method and takes on a single value $\alpha=1.5$. In
contrast, for large values of $a>10^{-2}$, the exponent $\alpha$
depends only on the order $\ell$ of the DFA method and takes on values
$\alpha=\ell+1$. This behavior is identical with the behavior obtained
for the exponential function in Fig.~\ref{trends3}(b). For intermediate
values of $a$, we observe a crossover in the scaling behavior of the
fluctuation function $F(n)$ from $\alpha=1.5$ to $\alpha=\ell+1$.

(iii) {\it Finally, we consider the general power-law function}:
$y=(x+a)^{\lambda}$, where $0<x\leq 1$, $x=i/N_{max}$, $i=1,...,
N_{max}$, $N_{max}=2^{17}$, the power $\lambda$ takes on real values
and the offset parameter $a$ is a positive constant. As in the case of
the logarithmic function, we find again that the DFA scaling exponent
$\alpha$ depends on the value of the offset parameter $a$
[Fig.~\ref{trends1}(a)]. For certain fixed values of $\lambda$ and with
increasing $a$, we observe a gradual transition in the fluctuation
function $F(n)$ from a scaling behavior spanning over a broad range of
scales $n$ characterized by a small value of the exponent $\alpha$ to
a crossover at intermediate scales $n$ for larger values of $a$, and
finally to a scaling spanning over all scales $n$ with exponent
$\alpha=3$ for large values of $a$ for DFA-2. In a previous
study~\cite{kunpre2001} we have found a specific relationship between
the DFA exponent $\alpha$ and the value of the power $\lambda$ for the
case of power-law function with offset parameter $a=0$:
$\alpha=\ell+1$ for $\lambda> \ell-0.5$; $\alpha\simeq \lambda+1.5$
for $-1.5<\lambda< \ell-0.5$; $\alpha=0$ for $\lambda<-1.5$, where
$\ell$ is the order of polynomial fit in the DFA-$\ell$ method. Our
current analysis shows that this behavior is even more complicated
when $a>0$ [Fig.~\ref{trends1}(b)]. Specifically, we find that for
values of $a<10^{-5}$ the scaling exponent $\alpha$ [obtained in the
fitting range $n \in (30,3000)$] depends only on the value of the
power $\lambda$: $\alpha\simeq \lambda+1.5$. In contrast, for large
values of the offset parameter $a>10^{-2}$, we find that the exponent
$\alpha$ depends only on the order $\ell$ of the DFA method, and takes
on values $\alpha=\ell+1$, which is similar to the results obtained for the
general exponential and logarithmic functions in this range of $a$
[Fig.~\ref{trends3}(b) and Fig.~\ref{trends2}(b)]). For intermediate
values of $a$ and for $-1.5 <\lambda< \ell-0.5$, we observe a crossover in
the scaling behavior of the fluctuation function $F(n)$ from
$\alpha\simeq \lambda+1.5$ to $\alpha=\ell+1$. Further, we find that
for $\lambda> \ell-0.5$, the DFA-$\ell$ scaling exponent remains
constant $\alpha=\ell+1$, and does not depend on the values of the
offset parameter $a$ --- we note that for $\lambda=0.41$ (close to
$\lambda=0.5=\ell-0.5$ for DFA-1) the dependence of $\alpha$ on $a$ is
close to a horizontal line [Fig.~\ref{trends1}(b)].

\smallskip
\noindent {\bf Analytic arguments}

Our results show that for large values of the offset parameter $a$,
the detrended fluctuation function $F(n)$ for all three analytic
functions --- exponential, logarithmic, and power-law --- exhibits
identical slope, where the DFA scaling exponent $\alpha$ does not
depend on the particular functional form but only the order $\ell$
of the DFA method: $\alpha=\ell+1$
[Fig.~\ref{trends3}(b),~\ref{trends2}(b) and~\ref{trends1}(b)]. The reason
for this common behavior is that (i) for large values of $a$, in each
DFA box of a given length $n$, all three functions can be expanded in
converging Taylor series, allowing for a perfect fit by a finite order
polynomial function, and (ii) that, due to this convergence, the same
polynomial function can be used when shrinking the box length $n$. In
contrast, for very small values of the offset parameter $a$, the DFA
results for all three functions are distinctly different and does not
depend on the order $\ell$ of the DFA method.  Below we give some
general analytic arguments for the dependence of the DFA exponent
$\alpha$ on the offset parameter $a$ presented in
Figs.~\ref{trends3},~\ref{trends2} and~\ref{trends1}.

\smallskip
\noindent (i) {\it General exponential function}: $y=\mbox{exp}(x+a), 0<x\leq 1$.    

 First, we substitute the variable $x$ by $z=x+a$: $y=e^z, z \in
(a,1+a]$. Next, we consider a DFA box starting at the coordinate
$z^{\prime}=s$ and ending at $z''=s+t$, where $t$ is
proportional to the number of points $n$ in the box ---
$t=(1+a-a)n/N_{max}=n/N_{max}$. For any value of $z \in (s,s+t)$ we
can expand the function in a Taylor series:

\begin{eqnarray}
e^z=\mbox{exp}(s+z_0)|_{0<z_0<t}
=e^s\left[1+z_0+\frac{z_0^2}{2!}+...\right].
\label{eqnq1}
\end{eqnarray}

Since this expansion converges, a finite polynomial function can
accurately approximate the exponential function in each DFA box. We note
that the DFA-$\ell$ method applied to above polynomial functions gives the 
scaling exponent $\alpha=\ell+1$ (see~\cite{kunpre2001}). Thus, for any
exponential function we find that the DFA scaling does not depend on
the value of the offset parameter $a$ and depends only on the order $\ell$ of
the polynomial fit in the DFA-$\ell$ procedure [Fig.~\ref{trends3}(b)].

\smallskip
\noindent (ii){\it General logarithmic  function}: $y=\mbox{log}_{10}(x+a), 0<x\leq 1$.  

First, we substitute the variable $x$ by $z=x+a$:
$y=\mbox{log}_{10}(z), z \in (a,1+a]$. Next, we consider a DFA box
starting at the coordinate $z^{\prime}=s$ and ending at
$z''=s+t$, where $t$ is proportional to the number of points
$n$ in the box --- $t=n/N_{max}$. For any value of $z \in (s,s+t)$ the
Taylor expansion is:

\begin{eqnarray}
\mbox{log}_{10}(z)&=&\mbox{log}_{10}(s+z_0)|_{0<z_0<t}\nonumber \\
&\sim& \mbox{ln}(1+z_0/s)\nonumber \\
&=&\frac{z_0}{s}-\frac{1}{2}\left (\frac{z_0}{s}\right )^2+...
\frac{(-1)^{m-1}}{m}\left (\frac{z_0}{s}\right )^m+....
\label{eqnq2}
\end{eqnarray}

This series is converging only when $z_0/s<1$, i.e.,
$z_0<s$. Since $z_0 \in (0,t)$, the condition for convergence in any
DFA box $(s,s+t)$ partitioning the function is $t<s$. From
$t=n/N_{max}$ and $s \in [a,1+a-t]$, we find that if $a>n/N_{max}$, the
logarithmic function in all DFA boxes is converging, and thus each box
can be approximated by a polynomial function, leading to scaling
exponent $\alpha=\ell+1$ --- depending only on the order $\ell$ of the
DFA-$\ell$ method [Fig.~\ref{trends2}].

When $t>s$, for certain values of $z_0 \in (0,t)$, the series in
Eq.~(\ref{eqnq2}) is diverging. Since $s \in [a,1+a-t]$, for
$s=a<t=n/N_{max}$, we find that the logarithmic function is divergent
in the first DFA box $(a,a+t)$, leading to deviation in the DFA
scaling for small values of $a$ [Fig.~\ref{trends2}].

\smallskip
\noindent (iii) {\it General power-law  function}: $y=(x+a)^{\lambda}, x\in(0,1]$.  

First, we substitute the variable $x$ by $z=x+a$:
$y=z^{\lambda}, z \in (a,1+a]$. Next, we consider a DFA box
starting at the coordinate $z^{\prime}=s$ and ending at
$z''=s+t$, where $t$ is proportional to the number of points
$n$ in the box --- $t=n/N_{max}$. For any value of $z \in (s,s+t)$ the
Taylor expansion is:

\begin{eqnarray}
z^{\lambda}&=&(s+z_0)^\lambda|_{0<z_0<t}\nonumber \\
&\sim& \left(1+\frac{z_0}{s}\right)^\lambda \nonumber \\
&=&1+\lambda\frac{z_0}{s}+\frac{\lambda(\lambda-1)}{2!}\left
(\frac{z_0}{s}\right)^2+....
\label{eqnq3}
\end{eqnarray}

Similar to the case of the logarithmic function, this series is converging
only when $z_0/s<1$. Following the same arguments as for the
logarithmic function we find that when $a>n/N_{max}$, the power-law
function is converging in any DFA box, and thus can be approximated by a
polynomial function, leading to the scaling exponent $\alpha=\ell+1$
[Fig.~\ref{trends1}], which is identical to the case of exponential and
logarithmic function.

In contrast, for $a<n/N_{max}$, the power-law function is divergent in
the first DFA box $(a,a+t)$, as in the case of the logarithmic
function, leading to a deviation in the scaling of $F(n)$ for small
values of $a$ [Fig.~\ref{trends1}]. While in the case of logarithmic
function this divergence leads to a fixed scaling exponent
$\alpha=1.5$, for power-law functions the value of the scaling
exponent $\alpha$ depends also on the power $\lambda$ [Fig.~\ref{trends1}].

We note that the above arguments can be used to estimate the results
of the DFA method for other functions. For all functions which can be
expanded in convergent Taylor expansion of a polynomial form in each
DFA box partitioning the function, the DFA method leads to identical
scaling results with the exponent $\alpha=\ell+1$, which is a notable
inherent limitation of the method. When there is divergent behavior in
some or all of the DFA boxes partitioning a function, the DFA scaling
exhibits crossover behavior to different values of the scaling
exponent $\alpha$ which depends on the functional form and the
specific parameters of the function.

\section{Conclusions}\label{Conclusion}

In summary, our study shows that linear transforms do not change the
scaling properties of a signal. However, the correlation properties of
a signal change after applying a polynomial filter. Moreover, such
change depends on the type of correlations (positive or
anti-correlations) in the signal, as well as on the power (odd or
even) of the polynomial filter. For the logarithmic filter we find
that the scaling behavior of the transformed signal remains unchanged
only when the original signal satisfies certain type of correlations
(characterized by scaling exponent within a given range). Comparing
the ``apparent'' scaling behavior of the exponential, logarithmic, and
power-law functions we find that within certain range for the values
of the parameters, the DFA fluctuation function $F(n)$ exhibits an
identical slope, and that the DFA results of a class of other analytic
functions can be reduced to these three cases. We attribute this
behavior to specific limitations of the DFA method. Therefore, careful
tests are necessary to accurately estimate the correlation properties
of signals after nonlinear transforms.

\section*{Acknowledgments} 
This work was supported by NIH Grant HL071972 and NIH/National Center
for Research Resources (Grant No. P41RR13622) and by the Spanish
Ministerio de Ciencia y Tecnologia (grants BFM2002-00183 and
BIO2002-04014-CO3-02).

\appendix

%{\bf DFA-1 in exponential functions}

\section{DFA-1 in exponential functions}

We consider an exponential function of the type $\exp (cx+a) $, where
the parameters $c$ and $a$ take on real values. The first step of the
DFA method is to integrate the signal [Sec.\ref{secpuren}]:

\begin{equation}
\int_{0}^{x}\exp (\frac{cy}{N}+a)dy=\allowbreak N\frac{e^{\frac{cx}{N}
+a}-e^{a}}{c}, 
\label{d2}
\end{equation}

where $N$ is the length of the signal and $x\in (0,N]$. We divide the
variable in the exponential by $N$, so that ($x/N$) is in the interval
$(0,1]$, as considered in Sec.~\ref{sectrends}. The next step of the
DFA method is to divide the integrated signal into boxes of length
$n$. For DFA-1, the squared detrended fluctuation function in the
$k-$th box, $F^{2}(n,k)$, is

\begin{equation}
F^{2}(n,k)=\frac{1}{n}\int_{(k-1)n}^{kn}\left[ N\frac{e^{\frac{cx}{N}
+a}-e^{a}}{c}-(b_{k}x-d_{k})\right] ^{2}dx, 
\label{d3}
\end{equation}

where the parameters $b_{k}$ and $d_{k}$ are obtained by a linear fit
to the integrated signal using least squares in the $k-$th box. These
two parameters can be obtained analytically, although their
expressions are too long. To obtain the squared detrended fluctuation
function for the entire signal partitioned in non-overlapping boxes of
length $n$, we sum over all boxes and calculate the average value:

\begin{widetext}
\begin{equation}
F^{2}(n)=\frac{1}{N/n}\sum_{k=1}^{N/n}F^{2}(n,k)=\frac{1}{N/n}
\sum_{k=1}^{N/n}\frac{1}{n}\int_{(k-1)n}^{kn}\left[ N\frac{e^{\frac{cx}{N}
+a}-e^{a}}{c}-(b_{k}x-d_{k})\right] ^{2}dx. 
\label{d4}
\end{equation}
\end{widetext}

Here, the index $k$ in the sum ranges from $1$ to $N/n$ (there are $
N/n$ boxes of length $n$ in the signal of length $N$). Using the
analytical expressions for $b_{k}$ and $d_{k}$, $F^{2}(n)$ can be
presented analytically in the form:

\begin{equation}
F^{2}(n)=g(n)\cdot h(n),
\label{d5}
\end{equation}

where
\begin{widetext}
\begin{equation}
g(n)=\left\{ -8Nc^{2}n^{2}\left( 1+e^{\frac{cn}{N}}+e^{\frac{2cn}{N}}\right)
+c^{3}n^{3}\left( e^{\frac{2cn}{N}}-1\right) +24N^{2}\left[ -\left( e^{\frac{
cn}{N}}-1\right) ^{2}N-cn+cne^{\frac{2cn}{N}}\right] \right\}  
\label{d6}
\end{equation}
\end{widetext}
and

\begin{equation}
h(n)=\frac{e^{2a}\left( e^{2c}-1\right) N^{2}}{2c^{6}\left( e^{\frac{2cn}{N}
}-1\right) n^{3}}. 
\label{d7}
\end{equation}

Due to the complexity of $g(n)$ and $h(n)$, the expression of $
F^{2}(n) $ is very complicated. However, as $n<N$ (and usually, $n\ll N$),
one can expand $F^{2}(n)\,$in powers of $n$ to obtain:

\begin{equation}
F^{2}(n)\simeq \allowbreak \frac{c\left( e^{2c}-1\right)e^{2a}}{1440N^{2}}
n^{4}. 
\label{d8}
\end{equation}

Finally, for the detrended fluctuation function $F(n)$ we obtain:

\begin{equation}
F(n)\simeq \allowbreak \sqrt{\frac{c\left( e^{2c}-1\right) }{1440}}
\frac{e^{a}}{N}n^{2}.
\label{d9}
\end{equation}

Thus the DFA-1 scaling exponent is $\alpha =2$ (in agreement with the
numerical simulation in Sec.~\ref{sectrends}, Fig.~\ref{trends3}). In
general, we can obtain in a similar way that $\alpha =\ell +1$, when
DFA-$\ell$ with an order $\ell$ of polynomial fit is used.

%\end{multicols}

\end{document}